\documentclass[12pt]{article}
\pdfoutput=1
\usepackage{amsmath,amsfonts,graphicx,color,bbm,tikz,bm}
\usepackage{comment}
\usepackage[nosort]{cite}
\usepackage{subfigure}
\usetikzlibrary{calc,positioning}
\usetikzlibrary{patterns,arrows,decorations.pathreplacing}
\usepackage{caption}
\usepackage{ulem}
\tikzset{>=stealth}
\usepackage{lipsum}
\usepackage{tabularx}
\usepackage{fullpage}
\usepackage{hyperref}
\usepackage{bbold}

\makeatletter\@addtoreset{equation}{section}\makeatother

\newcommand{\be}{\begin{equation}}
\newcommand{\ee}{\end{equation}}
\def\beq{\begin{equation}}
\def\eeq{\end{equation}}
\newcommand{\bea}{\begin{eqnarray}}
\newcommand{\eea}{\end{eqnarray}}
\newcommand{\Tr}{{\rm Tr\,}}

\newcommand{\ket}[1]{{\left| {#1} \right>}}

\renewcommand{\title}[1]{\vbox{\center\LARGE{#1}}\vspace{3mm}}
\renewcommand{\author}[1]{\vbox{\center{#1}}\vspace{3mm}}

\newcommand{\email}[1]{\vbox{\center\tt#1}\vspace{3mm}}

\begin{document}
\begin{titlepage}

\begin{center}
{\large {\bf Yang-Baxter solutions from commuting operators }}

\author{ Pramod Padmanabhan,$^a$ Kun Hao,$^{b,c}$
and Vladimir Korepin$^{d}$}

{$^a${\it School of Basic Sciences,\\ Indian Institute of Technology, Bhubaneswar, 752050, India}}
\vskip0.1cm
{$^b${\it Peng Huanwu Center for Fundamental Theory, Xi'an, 710127, China}  }
\vskip0.1cm
{$^c${\it Institute of Modern Physics, \\ Northwest University, Xi'an, 710127, China }}
\vskip0.1cm
{$^d${\it C. N. Yang Institute for Theoretical Physics, \\ Stony Brook University, New York 11794, USA}}

\email{pramod23phys@gmail.com, haoke72@163.com, vladimir.korepin@stonybrook.edu}

\vskip 0.5cm 

\end{center}


\abstract{
\noindent 
We construct $R$-matrices (with a multidimensional spectral parameter) that include  additive as well as non-additive parameters. They satisfy the colored Yang-Baxter equation. The solutions depend on a set of commuting operators. They change with the representation of the latter. The associated Yang-Baxter algebra and the spectrum of the transfer matrix are also studied.
}

\end{titlepage}

\section{Introduction}
\label{sec:Introduction}

Several theoretical tools have been developed to study the spectrum of quantum many-body systems. Early methods include the coordinate Bethe ansatz \cite{Bethe1931OnTT,2014bewa.book.....G}, which evolved into the {\it quantum inverse scattering method} (QISM) or the {\it algebraic Bethe ansatz} (ABA) \cite{Sklyanin1979QuantumIP,Sklyanin1982QuantumVO,Korepin1993QuantumIS,slavnov2018algebraic,Sogo1982QuantumIS}\footnote{For a more extensive list of references see reviews of the evolution of this technique, pedagogical accounts and collections of articles in \cite{Takhtajan2017ScientificHO,Baxter1982ExactlySM,Sutherland2004BeautifulM7}.}. The starting point of the ABA is the $R$-matrix. They appear in two forms - additive and non-additive. Additive solutions underlie well-known spin chains including the Heisenberg ($XXX$) \cite{Faddeev1984SpectrumAS}, the $XXZ$ \cite{Kirillov1986ExactSO}, the $XYZ$ spin chains \cite{Baxter1982ExactlySM,Takhtadzhan1979THEQM,Jimbo1990} and the Potts model \cite{Fateev1982SelfdualSO}. The eigenstates \cite{Burdik2014} of the former spin chains are in one-to-one correspondence with supersymmetric gauge theories \cite{Klimcik2002,Beisert2005,Ahn2008,NEKRASOV200991}. Systematic algorithms for finding 4 by 4 additive $R$-matrices include the boost operator and differential operator methods \cite{Leeuw2020YangBaxterAT,Vieira2017SolvingAC, Garkun2024NewSD}. These result in nearest-neighbor models. Recently integrable models with quasi-local interactions have appeared as integrable deformations of the $XXZ$ spin chain \cite{Gombor2021IntegrableSC,Pozsgay2021IntegrableHD, Gombor2022IntegrableDO}. Known non-additive $R$-matrices are fewer in number. Non-additive solutions give the chiral Potts model \cite{Baxter1988NewSO,AuYang2016About3Y,Bazhanov:1989nc, Perk2015TheEH}. Shastry's non-additive $R$-matrix gives the Hubbard model \cite{Shastry1986ExactIO} and proves its integrability \cite{Shiroishi1995YangBaxterEF,esslerfrahmkorepin2005}. Another  example of a non-additive solution depends  on spectral parameters belonging to the  $SL(2, \mathbb{C})$ group (See section VI.4 of \cite{Korepin1993QuantumIS}). A few 4 by 4 solutions can be found in \cite{Bracken1994SolutionsOT,Zhang2020NewRW,Babichenko2012MultiparametricRF,Corcoran2023AllR}. With the inclusion of additive parameters they solve the so called {\it colored} Yang-Baxter equation \cite{Ge:1992mi,Sun1995ClassificationOS,Wang1996ClassificationOE,Khachatryan2013OnTS,Luan1999ColoredSO}.

In this work we will construct non-additive solutions of the  Yang-Baxter equation and multi-parameter solutions of the colored Yang-Baxter equation [some $R$-matrices which we found have multi-dimensional spectral parameter]. Our method is purely algebraic. We show that for every set of Abelian operators there exists an entire family of non-additive and multi-parameter $R$-matrices acting on the qudit Hilbert space. One such choice is the set of mutually orthogonal projectors. They can be both Hermitian and non-Hermitian depending on whether the projectors are orthogonal or oblique, respectively. We describe the proeprties of such $R$-matrices in this paper. 

Our findings are organised as follows. A brief review of $R$-matrices and the Yang-Baxter equations is in Sec. \ref{sec:mparameterR}. The algebraic method to derive the non-additive and multi-parameter $R$-matrices is also included here. The Yang-Baxter algebra for the rank 1 qubit projector spin chain is in Sec. \ref{sec:YangBaxterAlgebra}. We end with a summary and future directions in Sec. \ref{sec:conclusion}. Non-additive solutions based on supersymmetry algebras are presented in Appendix \ref{app:SUSYnar}. This shows the broad scope of the general solutions introduced in Sec. \ref{sec:mparameterR}. The novelty of the non-additive solutions found in Sec. \ref{sec:mparameterR} is tested in Appendix \ref{app:inequivalence}. Here we show that these $R$-matrices are inequivalent to known non-additive $R$-matrices in the literature. Proofs related to the Yang Baxter algebra and transfer matrices can be found in Appendices \ref{app:relations_ABD} and \ref{app:transfermatrices} respectively.


\section{Multi-parameter $R$-matrices}
\label{sec:mparameterR}
The $R$-matrix $R_{j,j+1}(u)\equiv R_j(u)$, is an invertible operator acting on a Hilbert space $\mathcal{H}=\bigotimes\limits_{j=1}^N~\mathbb{C}^d_j$. It has non-trivial support on two consecutive sites $j$ and $j+1$. On the rest of the Hilbert space it acts as the identity operator. The complex number $u$ is called the spectral parameter. When this operator satisfies the Yang-Baxter equation (YBE)\footnote{The YBE also appears in an equivalent form $$R_{12}(u)R_{23}(u+v)R_{12}(v)=R_{23}(v)R_{12}(u+v)R_{23}(u),$$ in the literature. This is obtained by setting $j=1$ in \eqref{eq:YBEadditive}. We will use both these forms in this work.} of the form,
\begin{equation}\label{eq:YBEadditive}
    R_j(u)R_{j+1}(u+v)R_j(v) = R_{j+1}(v)R_j(u+v)R_{j+1}(u),
\end{equation}
the $R$-matrix is said to be additive in $u$. This is the braided form of the YBE as it resembles the braid relation, $\sigma_j\sigma_{j+1}\sigma_j=\sigma_{j+1}\sigma_j\sigma_{j+1}$. The operators ($\sigma_j$) generate the braid group\footnote{Given a braid representation, it is possible to construct spectral parameter dependent $R(u)$-matrices by a process known as {\it Baxterization} \cite{Jones1990}. In this perspective the braid relation is also known as the constant Yang-Baxter equation.}. Every such $R_j(u)$-matrix is in one-to-one correspondence with an invertible operator $$\check{R}_j(u)=R_j(u)P_j.$$ Here $P_j\equiv P_{j,j+1}$ is the permutation operator exchanging the states on sites $j$ and $j+1$. The $\check{R}_j(u)$ satisfies\footnote{This is the simplified version of the more general form $$ \check{R}_{j,j+1}(u-v)\check{R}_{j,j+2}(u)\check{R}_{j+1,j+2}(v) = \check{R}_{j+1,j+2}(v)\check{R}_{j,j+2}(u)\check{R}_{j,j+1}(u-v).$$}
\begin{equation}\label{eq:qYBEadditive}
    \check{R}_{12}(u-v)\check{R}_{13}(u)\check{R}_{23}(v) = \check{R}_{23}(v)\check{R}_{13}(u)\check{R}_{12}(u-v),
\end{equation}
known as the $RTT$ form of YBE. This version is useful for proving the integrability of spin chain systems (See \cite{Korepin1993QuantumIS,slavnov2018algebraic} for a review). The relation \eqref{eq:qYBEadditive} is obtained from the YBE \eqref{eq:YBEadditive} by using the defining relations of the permutation generators,
\begin{equation}
    P_j^2=P_j,~~P_jP_{j+1}P_j=P_{j+1}P_jP_{j+1},~~P_jP_k=P_kP_j,~|j-k|>1.
\end{equation}

The non-additive analogs of \eqref{eq:YBEadditive} and \eqref{eq:qYBEadditive} are,
\begin{equation}\label{eq:YBEnonadditive}
    R_j(\bm{\mu},\bm{\nu})R_{j+1}\bm{(\mu}, \bm{\omega})R_j(\bm{\nu}, \bm{\omega}) = R_{j+1}(\bm{\nu}, \bm{\omega})R_j(\bm{\mu}, \bm{\omega})R_{j+1}(\bm{\mu}, \bm{\nu}),
\end{equation}
and 
\begin{equation}\label{eq:qYBEnonadditive}
    \check{R}_{12}(\bm{\mu},\bm{\nu})\check{R}_{13}(\bm{\mu}, \bm{\omega})\check{R}_{23}(\bm{\nu}, \bm{\omega}) = \check{R}_{23}(\bm{\nu}, \bm{\omega})\check{R}_{13}(\bm{\mu}, \bm{\omega})\check{R}_{12}(\bm{\mu}, \bm{\nu}).
\end{equation}
The spectral parameters $(\bm{\mu}, \bm{\nu})$ are tuples $\bm{\mu}=\{\mu_1, \mu_2,\cdots \}$ and $\bm{\nu}=\{\nu_1, \nu_2,\cdots \}$ with the same cardinality. 

The multi-parameter $R$-matrices, $R(u;\bm{\mu},\bm{\nu})$ include both additive ($u, v, \cdots$) and non-additive ($\bm{\mu}, \bm{\nu},\cdots$) spectral parameters. They satisfy the {\it colored Yang-Baxter equation} (cYBE) \cite{Ge:1992mi,Sun1995ClassificationOS,Wang1996ClassificationOE},
\begin{equation}\label{eq:cYBE}
    R_j(u;\bm{\mu},\bm{\nu})R_{j+1}(u+v;\bm{\mu}, \bm{\omega})R_j(v;\bm{\nu}, \bm{\omega}) = R_{j+1}(v;\bm{\nu}, \bm{\omega})R_j(u+v;\bm{\mu}, \bm{\omega})R_{j+1}(u;\bm{\mu}, \bm{\nu}). 
\end{equation}
The RTT form of this equation reads,
\begin{equation}\label{eq:cqYBE}
    \check{R}_{12}(u-v;\bm{\mu},\bm{\nu})\check{R}_{13}(u;\bm{\mu}, \bm{\omega})\check{R}_{23}(v;\bm{\nu}, \bm{\omega}) = \check{R}_{23}(v;\bm{\nu}, \bm{\omega})\check{R}_{13}(u;\bm{\mu}, \bm{\omega})\check{R}_{12}(u-v;\bm{\mu}, \bm{\nu}).
\end{equation}

The primary use of the YBE is to show that the transfer matrices of certain spin chains commute with each other for different spectral parameters \cite{Korepin1993QuantumIS,slavnov2018algebraic}. 
The $R$-matrix is used to construct the transfer matrix defined as
\begin{equation}\label{eq:transferTI}
    \mathcal{T}(u;\bm{\mu}, \bm{\nu}) = \Tr_0\left[\check{R}_{0N}(u;\bm{\mu}, \bm{\nu})\cdots \check{R}_{02}(u;\bm{\mu}, \bm{\nu})\check{R}_{01}(u;\bm{\mu}, \bm{\nu})\right].
\end{equation}
The trace is taken over the auxiliary space, indexed by $0$, of the {\it monodromy matrix},
\begin{equation}\label{eq:monodromyTI}
    T(u;\bm{\mu}, \bm{\nu}) = \check{R}_{0N}(u;\bm{\mu}, \bm{\nu})\cdots \check{R}_{02}(u;\bm{\mu}, \bm{\nu})\check{R}_{01}(u;\bm{\mu}, \bm{\nu}).
\end{equation}

By virtue of \eqref{eq:cqYBE} we can show that the monodromy matrices satisfy the $RTT$-relation \cite{Korepin1993QuantumIS,slavnov2018algebraic},
\begin{equation}\label{eq:coloredRTT}
    \check{R}_{12}(u-v;\bm{\mu},\bm{\nu})T_{1}(u;\bm{\mu}, \bm{\omega})T_{2}(v;\bm{\nu}, \bm{\omega}) = T_{2}(v;\bm{\nu}, \bm{\omega})T_{1}(u;\bm{\mu}, \bm{\omega})\check{R}_{12}(u-v;\bm{\mu}, \bm{\nu}).
\end{equation}
The indices $1$ and $2$ index the auxiliary space. The entries of the monodromy matrices are operators. They act on the physical Hilbert space\footnote{This is also known as quantum space \cite{Korepin1993QuantumIS}.} of the system. By tracing over the auxiliary indices in \eqref{eq:coloredRTT} we find that 
\begin{equation}\label{eq:conservation}
    \left[\mathcal{T}(u;\bm{\mu}, \bm{\omega}), \mathcal{T}(v;\bm{\nu}, \bm{\omega}) \right] = 0.
\end{equation}
Thus the transfer matrices commute for different values of the spectral parameters. This introduces a tower of commuting operators when the transfer matrices are expanded in a Taylor series about specific values of the spectral parameter. The operators in this expansion can be both local and non-local. This expansion can be carried out in either of the additive or non-additive parameters or both. Most integrable models are studied for the additive case. When only the non-additive parameters are present then \eqref{eq:qYBEnonadditive} is used to construct the transfer matrices. The integrable models in this situation have received comparatively lesser attention \cite{Zhang2020NewRW}. Next we use algebraic methods to construct these multi-parameter $R$-matrices.



\subsection{Algebraic solutions to \eqref{eq:YBEnonadditive}}
\label{subsec:solutionsYBEnonadditive}
Consider a 2-qudit $R$-matrix in the factorised form, 
\begin{equation}\label{eq:generalRfactorised}
    \check{R}_{12}(\bm{\mu}, \bm{\nu})=Q_1(\bm{\mu})Q_2(\bm{\nu}).
\end{equation}
Here $Q(\bm{\mu})$ is any invertible operator\footnote{Note that this ansatz solves the non-additive YBE (\eqref{eq:qYBEnonadditive}) even if the invertibility condition for $Q(\bm{\mu})$ is not satisfied. However the resulting operators cannot be $R$-matrices which require invertibility.} on $\mathbb{C}^d$ depending on the parameter set $(\bm{\mu})$. It is easily seen to satisfy the non-additive YBE \eqref{eq:qYBEnonadditive} as both sides of this equation trivially simplifies to 
$$Q_1^2(\bm{\mu})Q_2^2(\bm{\nu})Q_3^2(\bm{\omega}).$$
It presents a non-trivial solution to the braided form of the non-additive YBE in \eqref{eq:YBEnonadditive}. This is due to the fact that the operators $Q_1(\bm{\mu})Q_2(\bm{\nu})P_{12}$ and $Q_2(\bm{\mu})Q_3(\bm{\omega})P_{23}$ do not commute with each other. Thus depending on the representation for the invertible operator $Q$, the $R$-matrices in \eqref{eq:generalRfactorised} provide a large class of solutions to the non-additive YBE \eqref{eq:qYBEnonadditive}.

This solution prompts a more general class of solutions to the non-additive YBE through the $R$-matrix :
\begin{equation}\label{eq:generalRnonfactorised}
    \check{R}_{12}(\bm{\mu}, \bm{\nu}) = \sum\limits_m~\left(Q_m(\bm{\mu})\right)_1\left(Q_m(\bm{\nu})\right)_2.
\end{equation}
The operators $Q_m(\bm{\mu})$, acting on on $\mathbb{C}^d$, are elements of a set of mutually commuting operators $$\left[Q_m(\bm{\mu}), Q_n(\bm{\mu})\right]=0.$$  They are indexed by the integers $m$, $n$. Additionally, note that $$\left[Q_m(\bm{\mu}), Q_n(\bm{\nu})\right]\neq 0,$$ when $\bm{\mu}\neq\bm{\nu}$. The identity operator on $\mathbb{C}^d$ belongs to this set. Note that the individual elements of this set need not be invertible. We only require that the non-additive $R$-matrix \eqref{eq:generalRnonfactorised} be invertible. The non-additive $R$-matrices of \eqref{eq:generalRnonfactorised} are elements of the corresponding universal enveloping algebras of these Abelian structures.  

Next we construct such Abelian structures on $\mathbb{C}^d$ that we will use in this work\footnote{We present a set of solutions based on supercharges in Appendix \ref{app:SUSYnar}.}. To this end consider the following non-additive $R$-matrix on $\mathbb{C}^d\otimes\mathbb{C}^d$,
\begin{equation}\label{eq:projectorNAR}
    \check{R}(\bm{\mu}, \bm{\nu}) = \mathbb{1} + \Pi_{\bm{\mu}}\otimes \Pi_{\bm{\nu}}.
\end{equation}
The $\Pi$'s are projectors on $\mathbb{C}^d$ parametrised by the tuples $\bm{\mu}$ and $\bm{\nu}$. The projectors can be either orthogonal (Hermitian) or oblique (non-Hermitian) depending on the number of parameters. While the projectors are not invertible, the combination with the identity operator makes it invertible. Its inverse is given by 
$$ \check{R}^{-1}(\bm{\mu}, \bm{\nu}) = \mathbb{1} -\frac{1}{2} \Pi_{\bm{\mu}}\otimes \Pi_{\bm{\nu}}.$$
This solution can be generalised. Consider the set of mutually orthogonal projectors on $\mathbb{C}^d$, $$\{\Pi^{(k)}_{\bm{\mu}}|k\in\{1,2,\cdots d\}\},$$ parametrised by $\bm{\mu}$. This is our desired example of an Abelian set. This gives the most general non-additive $R$-matrix\footnote{The inequivalence of this $R$-matrix to known non-additive $R$-matrices is shown in Appendix \ref{app:inequivalence}.}
\begin{equation}\label{eq:projectorNARgeneralqudit}
   \check{R}(\bm{\mu}, \bm{\nu}) = \mathbb{1} + \sum\limits_{k_1=1}^d\sum\limits_{k_2=1}^d~a_{k_1,k_2}~\Pi^{(k_1)}_{\bm{\mu}}\otimes \Pi^{(k_2)}_{\bm{\nu}}. 
\end{equation}
The coefficients $a_{k_1, k_2}$ are complex constants. The inverse of this operator is given by 
\begin{equation}\label{eq:projectorNARgeneralquditinverse}
   \check{R}^{-1}(\bm{\mu}, \bm{\nu}) = \mathbb{1} - \sum\limits_{k_1=1}^d\sum\limits_{k_2=1}^d~\frac{a_{k_1,k_2}}{1+a_{k_1,k_2}}~\Pi^{(k_1)}_{\bm{\mu}}\otimes \Pi^{(k_2)}_{\bm{\nu}}. 
\end{equation}
For certain values of the constants $a_{k_1, k_2}$ we obtain projectors on $\mathbb{C}^d\otimes\mathbb{C}^d$ of various ranks. To obtain a rank $r<d^2$ projector we set $r$ of the $d^2$ coefficients, $a_{k_1, k_2}$ to 1 and the rest to 0. There are $\binom{d^2}{r}$ rank $r$ projectors on $\mathbb{C}^d\otimes\mathbb{C}^d$. 

Since we focus on the qubit ($d=2$) case in the rest of the paper we will list their $R$-matrices for the different rank projectors below. In this case the non-additive $R$-matrix in \eqref{eq:projectorNARgeneralqudit} becomes 
\begin{equation}\label{eq:projectorNARgeneralqubit}
    \check{R}(\bm{\mu}, \bm{\nu}) = \mathbb{1} + a_{1,1}~\Pi^{(1)}_{\bm{\mu}}\otimes \Pi^{(1)}_{\bm{\nu}} + a_{1,2}~\Pi^{(1)}_{\bm{\mu}}\otimes \Pi^{(2)}_{\bm{\nu}} + a_{2,1}~\Pi^{(2)}_{\bm{\mu}}\otimes \Pi^{(1)}_{\bm{\nu}} + a_{2,2}~\Pi^{(2)}_{\bm{\mu}}\otimes \Pi^{(2)}_{\bm{\nu}}.
\end{equation}
Here $a_{j,k}$ are complex numbers. As the qubit case consists of just two orthogonal projectors we will use the notation $\Pi^{(1)}\equiv\Pi$ and $\Pi^{(2)}\equiv\Pi^\perp$, in what follows.

\paragraph{{\bf Rank 1} :}

In this case the $R$-matrices are obtained when just one of the four coefficients $a_{k_1, k_2}=1$ in \eqref{eq:projectorNARgeneralqubit}. There are four such operators. An example is 
\begin{equation}
    \check{R}(\bm{\mu}, \bm{\nu})  =   \mathbb{1} + \Pi_{\bm{\mu}}\otimes \Pi_{\bm{\nu}}. \label{eq:rank1NAR-1}
\end{equation}
 
\paragraph{{\bf Rank 2} :}

For this case two of the four coefficients are 1 and the rest are 0. 
There are 6 such $R$-matrices. Two of these act non-trivially on $\mathbb{C}^d\otimes\mathbb{C}^d$. These are 
\begin{eqnarray}
    \check{R}(\bm{\mu}, \bm{\nu}) & = &  \mathbb{1} + \Pi_{\bm{\mu}}\otimes \Pi_{\bm{\nu}} + \Pi^\perp_{\bm{\mu}}\otimes \Pi^\perp_{\bm{\nu}}, \label{eq:rank2NAR-3}\\
    \check{R}(\bm{\mu}, \bm{\nu}) & = &  \mathbb{1} + \Pi_{\bm{\mu}}\otimes \Pi^\perp_{\bm{\nu}} + \Pi^\perp_{\bm{\mu}}\otimes \Pi_{\bm{\nu}}. \label{eq:rank2NAR-4}
\end{eqnarray}
The remaining four act non-trivially on one of the two $\mathbb{C}^d$'s in the tensor product. An example is 
\begin{equation}
    \check{R}(\bm{\mu}, \bm{\nu})  =   \mathbb{1} + \Pi_{\bm{\mu}}\otimes \Pi_{\bm{\nu}} + \Pi_{\bm{\mu}}\otimes \Pi^\perp_{\bm{\nu}}=\mathbb{1} + \Pi_{\bm{\mu}}. \label{eq:rank2NAR-1}
\end{equation}


\paragraph{{\bf Rank 3} :}

Now three coefficients are 1 and the remaining is 0. There are four such non-additive $R$-matrices. An example is  
\begin{equation}
    \check{R}(\bm{\mu}, \bm{\nu})  =  \mathbb{1} + \Pi_{\bm{\mu}}\otimes \Pi_{\bm{\nu}} + \Pi_{\bm{\mu}}\otimes \Pi^\perp_{\bm{\nu}} + \Pi^\perp_{\bm{\mu}}\otimes \Pi_{\bm{\nu}}. \label{eq:rank3NAR-1}
\end{equation}


\paragraph{{\bf Rank 4} :}

In this case the $R$-matrix is just the identity operator as 
$$ \Pi_{\bm{\mu}}\otimes \Pi_{\bm{\nu}} + \Pi_{\bm{\mu}}\otimes \Pi^\perp_{\bm{\nu}} + \Pi^\perp_{\bm{\mu}}\otimes \Pi_{\bm{\nu}} + \Pi^\perp_{\bm{\mu}}\otimes \Pi^\perp_{\bm{\nu}} = \mathbb{1}. $$

For the above $R$-matrices one of the parameters from the tuple $\bm{\mu}$ can be taken as the spectral parameter. We can also include an additional additive spectral parameter. Such operators will solve the cYBE of \eqref{eq:cqYBE}. We will now illustrate this construction.

\subsection{Algebraic solutions to \eqref{eq:cqYBE}}
\label{subsec:solutionscqYBE}
An additive spectral parameter $u$ can be included in the non-additive $R$-matrices of \eqref{eq:projectorNARgeneralqudit} as
\begin{equation}\label{eq:projectormultiRqudit}
    \check{R}(u;\bm{\mu}, \bm{\nu}) = \mathbb{1} + \alpha u~\sum\limits_{k_1=1}^d\sum\limits_{k_2=1}^d~a_{k_1,k_2}~\Pi^{(k_1)}_{\bm{\mu}}\otimes \Pi^{(k_2)}_{\bm{\nu}}.
\end{equation}
Here $\alpha$ is an arbitrary constant. This solves \eqref{eq:cqYBE} with the inverse given by 
\begin{equation}\label{eq:projectormultiRquditinverse}
    \check{R}^{-1}(u;\bm{\mu}, \bm{\nu}) = \mathbb{1} - \alpha u~\sum\limits_{k_1=1}^d\sum\limits_{k_2=1}^d~\frac{a_{k_1,k_2}}{1+\alpha ua_{k_1,k_2}}~\Pi^{(k_1)}_{\bm{\mu}}\otimes \Pi^{(k_2)}_{\bm{\nu}}.
\end{equation}
We will adapt these solutions to the qubit case. Consider the following orthonormal basis for $\mathbb{C}^2$,
\begin{eqnarray}
    \ket{0_\mu} & = & \frac{1}{\sqrt{1+\mu^2}}\left[-\mu\ket{0}+\ket{1}\right]\equiv \frac{1}{\sqrt{1+\mu^2}}\begin{pmatrix}
        -\mu \\ 1
    \end{pmatrix}, \nonumber \\
    \ket{1_\mu} & = & \frac{1}{\sqrt{1+\mu^2}}\left[\ket{0}+\mu\ket{1}\right]\equiv \frac{1}{\sqrt{1+\mu^2}}\begin{pmatrix}
        1 \\ \mu
    \end{pmatrix},
\end{eqnarray}
parametrised by $\mu\in\mathbb{R}$. Here $\{\ket{0}, \ket{1}\}$ is the canonical basis of $\mathbb{C}^2$. Then the projector to a general product state $\ket{\psi}\equiv\ket{1_\mu}\otimes\ket{1_\nu}\in\mathbb{C}^2\otimes\mathbb{C}^2$ takes the form,
\begin{equation}\label{eq:PimuPinumatrices}
    \Pi_{\mu\mu}\otimes\Pi_{\nu\nu} = \frac{1}{(1+\mu^2)(1+\nu^2)}\begin{pmatrix}
        1 & \mu \\ \mu & \mu^2
    \end{pmatrix}\otimes \begin{pmatrix}
        1 & \nu \\ \nu & \nu^2
    \end{pmatrix} ; ~~ \mu,\nu\in\mathbb{R}.
\end{equation}
The individual projectors satisfy the relations,
\begin{equation}\label{eq:PimuPinurelations}
    \Pi_{\mu\mu}^2 = \Pi_{\mu\mu},~~ \Pi_{\nu\nu}^2 = \Pi_{\nu\nu};~~\left[\Pi_{\mu\mu}, \Pi_{\nu\nu}\right] = -\mathrm{i}\frac{(\mu-\nu)(1+\mu\nu)}{(1+\mu^2)(1+\nu^2)} Y,
\end{equation}
for different parameters $\mu$ and $\nu$. Here $Y$ is the second Pauli matrix. They have eigenvalues, $\{0, 1 \}$. These are the most general orthogonal projectors on $\mathbb{C}^2$ and are Hermitian. Clearly they do not commute with each other except when $\mu=\nu$ or when $\mu\nu=-1$. In the former case the two projectors coincide on $\mathbb{C}^2$. In the latter case they project to orthogonal subspaces of $\mathbb{C}^2$. The orthogonal projector is given by 
\begin{equation}
    \Pi_{\mu\mu}^\perp = \frac{1}{1+\mu^2}\begin{pmatrix}
        \mu^2 & -\mu \\ -\mu & 1
    \end{pmatrix}.
\end{equation}
Furthermore we have the relations,
\begin{equation}\label{eq:PiTL}
    \Pi_{\mu\mu}\Pi_{\nu\nu}\Pi_{\mu\mu} = \frac{(1+\mu\nu)^2}{(1+\mu^2)(1+\nu^2)}~\Pi_{\mu\mu} ,~~ \Pi_{\nu\nu}\Pi_{\mu\mu}\Pi_{\nu\nu} = \frac{(1+\mu\nu)^2}{(1+\mu^2)(1+\nu^2)}~\Pi_{\nu\nu}.
\end{equation}
This implies that the algebra generated by $\Pi_{\mu\mu}, \Pi_{\nu\nu}$ on $\mathbb{C}^2$ is spanned by $$\{\Pi_{\mu\mu}, \Pi_{\nu\nu}, \Pi_{\mu\mu}\Pi_{\nu\nu}, \Pi_{\nu\nu}\Pi_{\mu\mu}\}.$$ Note here that the operators $\Pi_{\mu\mu}\Pi_{\nu\nu}$ and $\Pi_{\nu\nu}\Pi_{\mu\mu}$ are non-Hermitian. So we can expand the set of projectors on $\mathbb{C}^2$ by including non-Hermitian projectors. Consider the set, 
\begin{equation}\label{eq:Pialgebra}
  \mathcal{P}=\left\{\Pi_{\mu_1\mu_2} =  \frac{1}{1+\mu_1\mu_2}\begin{pmatrix}
        1 & \mu_1 \\ \mu_2 & \mu_1\mu_2
    \end{pmatrix}\Bigg|\mu_1,\mu_2\in\mathbb{R}\right\}.
\end{equation}
When $\mu_1\neq \mu_2$ ($\mu_1=\mu_2$) these projectors are oblique (orthogonal). The operator algebra $\mathcal{P}$ is closed under matrix multiplication. They satisfy the relations,
\begin{equation}\label{eq:PiRelations}
    \Pi_{\mu_1\nu_1}\Pi_{\mu_2\nu_2} = \frac{(1+\mu_1\nu_2)(1+\mu_2\nu_1)}{(1+\mu_1\nu_1)(1+\mu_2\nu_2)}\Pi_{\mu_2\nu_1},~~ \Pi_{\mu_2\nu_2}\Pi_{\mu_1\nu_1} = \frac{(1+\mu_1\nu_2)(1+\mu_2\nu_1)}{(1+\mu_1\nu_1)(1+\mu_2\nu_2)}\Pi_{\mu_1\nu_2}.
\end{equation}
As the elements are non-linear functions of their parameters, this algebra is not closed under matrix addition. The algebra is infinite dimensional. The full identity is not an element but the partial identities, $\Pi_{00}$, $\Pi^\perp_{00}$ are included. The center of this algebra is trivial. 

The Hermitian projectors in this algebra are used to construct the multi-parameter $R$-matrices,
\begin{eqnarray}
    \check{R}_j(u;\mu,\nu) & = & \mathbb{1} + \alpha u\left[a_1\left(\Pi_{\mu\mu}\right)_{j}\left(\Pi_{\nu\nu}\right)_{j+1} + a_2\left(\Pi^\perp_{\mu\mu}\right)_{j}\left(\Pi_{\nu\nu}\right)_{j+1} \right. \nonumber \\ &  + & \left. a_3\left(\Pi_{\mu\mu}\right)_{j}\left(\Pi^\perp_{\nu\nu}\right)_{j+1} + a_4\left(\Pi^\perp_{\mu\mu}\right)_{j}\left(\Pi^\perp_{\nu\nu}\right)_{j+1}\right].
\end{eqnarray}

\section{Yang-Baxter algebra}
\label{sec:YangBaxterAlgebra}
Considering the monodromy matrix \eqref{eq:monodromyTI}, it is a $2\times2$ matrix in the auxiliary space $0$
\begin{equation}\label{eq:monodromy}
    T(u;{\mu}, {\nu}) = \check{R}_{0N}(u;{\mu}, {\nu})\cdots \check{R}_{02}(u;{\mu}, {\nu})\check{R}_{01}(u;{\mu}, {\nu})=
    \begin{pmatrix} \mathcal{A}(u;{\mu}, {\nu}) & \mathcal{B}(u;{\mu}, {\nu}) \cr \mathcal{C}(u;{\mu}, {\nu}) & \mathcal{D}(u;{\mu}, {\nu}) \end{pmatrix}_0.
\end{equation}
The elements are the operators $\mathcal{A}(u;{\mu}, {\nu})$ , $\mathcal{B}(u;{\mu}, {\nu})$, $\mathcal{C}(u;{\mu}, {\nu})$, and $\mathcal{D}(u;{\mu}, {\nu})$ acting on the whole quantum space $\bigotimes\limits_{j=1}^N~\mathbb{C}^d_j$.
Here we adopt the $R$-matrix of the following form based on \eqref{eq:projectormultiRqudit} to illustrate the algebraic structure of the current model
\begin{equation}\label{eq:R-matrix}
\check{R}_{j,j+1}(u;\mu,\nu) = \mathbb{1} +  u\left[\left(\Pi_{\mu\mu}\right)_{j}\left(\Pi_{\nu\nu}\right)_{j+1}\right]=\mathbb{1} + \frac{u}{(1+\mu^2)(1+\nu^2)}\begin{pmatrix}
        1 & \mu \\ \mu & \mu^2
    \end{pmatrix}_j \begin{pmatrix}
        1 & \nu \\ \nu & \nu^2
    \end{pmatrix}_{j+1}.
\end{equation}
Here $u$ is an additive spectral parameter. This $R$-matrix satisfy the Yang-Baxter equation, and possesses the following properties:
\begin{eqnarray}
\mbox{Initial condition:}& \check{R}_{12}(0;\mu,\nu)= \mathbb{1},\label{eq:Initial-R}\\
\mbox{PT-symmetry:}& \check{R}_{12}(u;\mu,\nu)=\check{R}_{12}(u;\mu,\nu)^{t_l}=\check{R}_{12}(u;\mu,\nu)^{t_1t_2}\neq\check{R}_{21}(u;\mu,\nu), \;l=1,2~\\
\mbox{Unitarity:} & \check{R}_{12}(u;\mu,\nu)\check{R}_{12}(\frac{-u}{u+1};\mu,\nu)=\mathbb{1},\label{eq:Unitarity}
\end{eqnarray}
The partial transpose of the $R$-matrix on either the $1$st quantum space or the $2$nd quantum space reverts to itself. This fact establishes the equivalence between the crossing unitarity\footnote{Regarding the additive parameter in the $R$-matrix, it is the ratio between the identity component and the projector component that primarily influences the outcomes. Consequently, the $R$-matrix can be reparameterized as $\check{R}_{12}(u)=\check{R}_{12}(u; \mu, \nu) =  \mathbb{1}u +  \eta\left[\left(\Pi_{\mu\mu}\right)_{1}\left(\Pi_{\nu\nu}\right)_{2}\right]$. Under this reparameterization, both unitarity and crossing unitarity can be simply expressed as $\check{R}_{12}(u)\check{R}_{12}(-u-\eta)=-u(u+\eta)\mathbb{1}$. This result can be compared with other integrable models, such as the Heisenberg XXX spin-$1\over2$ chain.} and the unitarity relation \eqref{eq:Unitarity} of the $R$-matrix.

For most integrable models, a set of algebraic relations among these four operators is encoded in the fundamental $RTT$-relation \eqref{eq:coloredRTT}
\begin{equation}
    \check{R}_{12}(u-v;{\mu},{\nu})T_{1}(u;{\mu}, {\omega})T_{2}(v;{\nu}, {\omega}) = T_{2}(v;{\nu}, {\omega})T_{1}(u;{\mu}, {\omega})\check{R}_{12}(u-v;{\mu}, {\nu}).
\end{equation}
As for the current model, the $RTT$-relation can only give rise to some long equations with the four operators mixed together. It is hard to get concise algebra relations from them.

Based on the definition of the above monodromy matrix \eqref{eq:monodromy} and $R$-matrix \eqref{eq:R-matrix}, we can verify that the operators $\mathcal{A}$, $\mathcal{B}$, $\mathcal{C}$, and $\mathcal{D}$ commute with themselves for different values of $u$ and $\mu$, but not for $\nu$ (the third parameter in monodromy matrix and the operators)
\begin{eqnarray}
&&[\mathcal{A}(u;\mu_1,\nu),\ \mathcal{A}(v;\mu_2,\nu)] =  [\mathcal{B}(u;\mu_1,\nu),\ \mathcal{B}(v;\mu_2,\nu)] =
 [\mathcal{C}(u;\mu_1,\nu),\ \mathcal{C}(v;\mu_2,\nu)] \nonumber\\
&& = [\mathcal{D}(u;\mu_1,\nu),\ \mathcal{D}(v;\mu_2,\nu)] = 0 .
\end{eqnarray}
The structure of operator $\mathcal{B}$ is exactly the same (isomorphism) as that of operator $\mathcal{C}$,
\begin{eqnarray}
\mathcal{B}(u;\mu,\nu) =
 \mathcal{C}(u;\mu,\nu),\quad
[\mathcal{B}(u;\mu_1,\nu),  \mathcal{C}(v;\mu_2,\nu)]=0 .
\label{relation_B-C}\end{eqnarray}
The operators $\mathcal{B}$ and $\mathcal{D}$ can also be related to the operator $\mathcal{A}$ by the following relations
\begin{eqnarray}
\mathcal{A}(u;\mu,\nu)\mu-\mathcal{B}(u;\mu,\nu) & = & \mu\,\mathbb{1}\label{relation_A-B},\\
\mathcal{A}(u;\mu,\nu)\mu^2-\mathcal{D}(u;\mu,\nu) & = & (\mu^2 - 1) \,\mathbb{1} .\label{relation_A-D}
\end{eqnarray}
We give the proof for the above relations \eqref{relation_B-C}, \eqref{relation_A-B} and \eqref{relation_A-D} in Appendix \ref{app:relations_ABD}.
Based on \eqref{relation_A-B} and \eqref{relation_A-D}, it is also easy to obtain the relation between the operators $\mathcal{B}$ and $\mathcal{D}$
\begin{eqnarray}
\mathcal{D}(u;\mu,\nu)-\mathcal{B}(u;\mu,\nu)\mu  =  \mathbb{1}.
\end{eqnarray}
The above relations will lead to the fact that all the operators $\mathcal{A}$, $\mathcal{B}$, $\mathcal{C}$, and $\mathcal{D}$ commute with each other (Abelian algebra) for different values of $u$ and $\mu$
\begin{eqnarray}
[\mathcal{A}(u;\mu_1,\nu),  \mathcal{B}(v;\mu_2,\nu)]=[\mathcal{A}(u;\mu_1,\nu),  \mathcal{C}(v;\mu_2,\nu)]=[\mathcal{A}(u;\mu_1,\nu),  \mathcal{D}(v;\mu_2,\nu)]\nonumber\\
=[\mathcal{B}(u;\mu_1,\nu),  \mathcal{C}(v;\mu_2,\nu)]=[\mathcal{B}(u;\mu_1,\nu),  \mathcal{D}(v;\mu_2,\nu)]=[\mathcal{C}(u;\mu_1,\nu),  \mathcal{D}(v;\mu_2,\nu)]=0.
\end{eqnarray}
This implies that the conservation law \eqref{eq:conservation} can be constructed using various combinations of the four operators.

In the following, we give the general expressions for the four operators based on their recursive relations. For brevity, we denote the operators $\mathcal{A}(u;{\mu}, {\nu})$ , $\mathcal{B}(u;{\mu}, {\nu})$, $\mathcal{C}(u;{\mu}, {\nu})$, and $\mathcal{D}(u;{\mu}, {\nu})$ among the monodromy matrix of total system sites $N$ as $\mathcal{A}_N$ , $\mathcal{B}_N$, $\mathcal{C}_N$, and $\mathcal{D}_N$, respectively.
Based on the relations (\ref{eq:recursive-A},$\sim$\ref{eq:recursive-D}) in Appendix \ref{app:relations_ABD} and also the relations \eqref{relation_B-C}, \eqref{relation_A-B} and \eqref{relation_A-D}, the operators have the following recursive relations in terms of themselves with one less system site $N-1$:
\begin{eqnarray}
\mathcal{A}_{N}&=&\left(\mathbb{1}+\frac{u}{1+\mu^2}(\Pi_{\nu\nu})_N+\frac{u\mu^2}{1+\mu^2}(\Pi_{\nu\nu})_N\right)\mathcal{A}_{N-1}-\frac{u\mu^2}{1+\mu^2}(\Pi_{\nu\nu})_N,\\
\mathcal{B}_{N}&=&\left(\mathbb{1}+\frac{u}{1+\mu^2}(\Pi_{\nu\nu})_N+\frac{u\mu^2}{1+\mu^2}(\Pi_{\nu\nu})_N\right)\mathcal{B}_{N-1}+\frac{u\mu}{1+\mu^2}(\Pi_{\nu\nu})_N,\\
\mathcal{C}_{N}&=&\left(\mathbb{1}+\frac{u}{1+\mu^2}(\Pi_{\nu\nu})_N+\frac{u\mu^2}{1+\mu^2}(\Pi_{\nu\nu})_N\right)\mathcal{C}_{N-1}
+\frac{u\mu}{1+\mu^2}(\Pi_{\nu\nu})_N,\\
\mathcal{D}_{N}&=&\left(\mathbb{1}+\frac{u}{1+\mu^2}(\Pi_{\nu\nu})_N+\frac{u\mu^2}{1+\mu^2}(\Pi_{\nu\nu})_N\right)\mathcal{D}_{N-1}
-\frac{u}{1+\mu^2}(\Pi_{\nu\nu})_N,
\end{eqnarray}
with $\mathcal{A}_1=\mathbb{1}+\frac{u}{1+\mu^2} (\Pi_{\nu\nu})_1$, $\mathcal{B}_1=\mathcal{C}_1=\frac{u\mu}{1+\mu^2} (\Pi_{\nu\nu})_1$, and $\mathcal{D}_1=\mathbb{1}+\frac{u\mu^2}{1+\mu^2} (\Pi_{\nu\nu})_1$ .

Let us further define
\begin{eqnarray}
G_j=\mathbb{1}+\frac{u}{1+\mu^2}(\Pi_{\nu\nu})_j+\frac{u\mu^2}{1+\mu^2}(\Pi_{\nu\nu})_j,
\end{eqnarray}
and
\begin{eqnarray}
\widetilde{G}_j=\frac{u}{1+\mu^2}(\Pi_{\nu\nu})_j.
\end{eqnarray}
Then we have the general expressions for the operators  $\mathcal{A}_N$ , $\mathcal{B}_N$, $\mathcal{C}_N$, and $\mathcal{D}_N$ with arbitrary system sites $N$.
\begin{eqnarray}
\mathcal{A}_N&=&\mathcal{A}_1\prod_{j=2}^N(G_j)-\sum_{j=2}^N\mu^2\widetilde{G}_j\prod_{k=j+1}^N(G_k),\\
\mathcal{B}_N&=&\mathcal{B}_1\prod_{j=2}^N(G_j)+\sum_{j=2}^N\mu\widetilde{G}_j\prod_{k=j+1}^N(G_k)=\mathcal{C}_N,\\
\mathcal{D}_N&=&\mathcal{D}_1\prod_{j=2}^N(G_j)-\sum_{j=2}^N\widetilde{G}_j\prod_{k=j+1}^N(G_k).
\end{eqnarray}
The derivations in this section can be generalized to more complicated cases ($\check{R}$ matrices) mentioned in previous sections.

As a result of the Abelian algebra obeyed by the $\mathcal{A}$, $\mathcal{B}$, $\mathcal{C}$ and $\mathcal{D}$ operators, the transfer matrix can be written down for an arbitrary number of sites\footnote{For more details see Appendix \ref{app:transfermatrices}.}. The expression depends on whether this length is even or odd. For a chain of length $2N$ we have,
\begin{equation}\label{eq:transferMatrixeven}
    \mathcal{T}(u;\mu,\nu) = 2\mathop{\sum\limits_{i_1,\cdots,i_{2k}=1}^{2N}}_{\{i_1\neq\cdots\neq i_{2k}\}}\sum\limits_{k=0}^{N}~\left(\beta^2+\gamma^2 \right)^{N-k}\alpha_{i_1}\cdots\alpha_{i_{2k}}\mathop{\prod\limits_{j=1}^{2N}}_{j\neq\{i_1,\cdots,i_{2k}\}}~\left(\Pi_{\nu\nu}\right)_j,
\end{equation}
where 
\begin{equation}
    \alpha_j  =  \mathbb{1} + \frac{u}{2}\left(\Pi_{\nu\nu}\right)_j,~~
    \beta  = \frac{l(1-\mu^2)}{2},~~\gamma = \mu l,~;~l=\frac{u}{1+\mu^2}
\end{equation}
A similar expression can be written down for the chain with odd length $2N+1$ as
\begin{equation}\label{eq:transferMatrixodd}
    \mathcal{T}(u;\mu,\nu) = 2\mathop{\sum\limits_{i_1,\cdots,i_{2k+1}=1}^{2N+1}}_{\{i_1\neq\cdots\neq i_{2k+1}\}}\sum\limits_{k=0}^{N}~\left(\beta^2+\gamma^2 \right)^{N-k}\alpha_{i_1}\cdots\alpha_{i_{2k+1}}\mathop{\prod\limits_{j=1}^{2N+1}}_{j\neq\{i_1,\cdots,i_{2k+1}\}}~\left(\Pi_{\nu\nu}\right)_j.
\end{equation}
The transfer matrices in both cases are a sum of commuting terms and hence its spectrum is also easily obtained without the machinery of the algebraic Bethe ansatz. The eigenstates are product states with every site filled by either $\ket{0_\nu}$ or $\ket{1_\nu}$. The eigenvalues can be read off from the expressions for the transfer matrices \eqref{eq:transferMatrixeven} and \eqref{eq:transferMatrixodd}, depending on the sequence of $\ket{0_\nu}$ and $\ket{1_\nu}$'s in the eigenstate. 
The spectrum is summarised in Table \ref{tab:spectrumtransferMatrix}.

\begin{table}[h!]
\centering
\begin{tabular}{ |c|c|c| } 
 \hline
 Eigenstate & Eigenvalue & Degeneracy \\
 \hline
 $\ket{\left(0_\nu\right)_1\cdots \left(0_\nu\right)_N}$ & 2 & 1 \\ 
 \hline
$\ket{\left(0_\nu\right)_1\left(1_\nu\right)_{j_1}\cdots\left(1_\nu\right)_{j_{2f+1}}\left(0_\nu\right)_N}$ & $2\left(1 + \frac{u}{2}\right)^{2f+1}$  & $\binom{N}{2f+1}$ \\ 
 \hline
$\ket{\left(0_\nu\right)_1\left(1_\nu\right)_{j_1}\cdots\left(1_\nu\right)_{j_{2f}}\left(0_\nu\right)_N}$ & $2\left(1 + \frac{u}{2}\right)^{2f} + 2\left(\frac{u}{2}\right)^{2f}$ & $\binom{N}{2f}$ \\ 
 \hline
 $\ket{\left(1_\nu\right)_1\cdots \left(1_\nu\right)_N}$ & $2\sum\limits_{k=0}^{\lfloor\frac{N-1}{2}\rfloor}~\binom{N}{k}\left(1 + \frac{u}{2}\right)^{k}\left(\frac{u}{2}\right)^{2(N-k)}$ & 1 \\
 \hline
 \end{tabular}
 \caption{The spectrum of the transfer matrices \eqref{eq:transferMatrixodd} and \eqref{eq:transferMatrixeven}. The spectrum varies for an odd ($2f+1$) or even ($2f$) number of $1_\nu$'s in the product states. Here $f\in\{0,1,\cdots,\lfloor \frac{N-1}{2}\rfloor\}$ ($f\in\{0,1,\cdots,\lceil \frac{N-1}{2}\rceil\}$) for odd (even) number of $1_\nu$'s. See Appendix \ref{app:transfermatrices} for more details.}
 \label{tab:spectrumtransferMatrix}
\end{table}

Since now, in our case, the initial condition \eqref{eq:Initial-R}of the $R$-matrix is not proportional to the permutation operator,
we can not get Hamiltonian with nearest neighbor interactions from the first order derivative of the transfer matrix.
The first order terms of the transfer matrix are just sum of projectors on one lattice site.

The second order terms of transfer matrix will give rise to a Hamiltonian with long range interactions (non-local). Each lattice site will interact with all other sites.
\begin{eqnarray}
H = \sum_{i,j\atop i\neq j}^N\left(\Pi_{\nu\nu}\right)_{i}\left(\Pi_{\nu\nu}\right)_{j}
\end{eqnarray}
There are $\binom{N}{2}$ terms in the above summation in total.
This is just like the interactions in the Gaudin model, but it does not have the denominators as that in the Gaudin model.
The spectrum of this Hamiltonian can be read off from the second order derivative of the transfer matrix spectrum.

\subsection*{Special cases}
\label{subsec:specialcases}
The $R$-matrix in \eqref{eq:R-matrix} turns into a projector for particular values of the additive spectral parameter $u$ (See Table \ref{tab:specialcases}). 
\begin{table}[h!]
\centering
\begin{tabular}{ |c|c|c| } 
 \hline
 $u$ & $R$-matrix & Rank   \\
 \hline
 $0$ &  $\mathbb{1}$ & 4\\ 
 \hline
$1$ & $\mathbb{1}-\left(\Pi_{\mu\mu}\right)\otimes\left(\Pi_{\nu\nu}\right)$ & 3   \\ 
 \hline
$\infty$ & $\left(\Pi_{\mu\mu}\right)\otimes\left(\Pi_{\nu\nu}\right)$ & 1  \\ 
 \hline
\end{tabular}
 \caption{The $R$-matrix \eqref{eq:R-matrix} reduces to a projector at certain values of $u$.}
 \label{tab:specialcases}
\end{table}

We think that these can be used to describe the center of the Yang-Baxter algebra. The special cases also can be useful for description of fusion.



\section{Conclusion}
\label{sec:conclusion}
This work introduces a new class of solutions to the colored Yang-Baxter equation, \eqref{eq:cqYBE}. The multi-parameter $R$-matrices include both additive ($u$) and non-additive ($\bm{\mu}$) parameters. The method used to obtain them is purely algebraic. As a result the solutions are on qudit spaces. The main result is that for every set of commuting operators there exists a family of such multi-parameter $R$-matrices \eqref{eq:generalRnonfactorised}. In particular, when the Abelian set of operators are a mutually orthogonal set of projectors the solution becomes,
$$ \check{R}(u;\bm{\mu}, \bm{\nu}) = 1 + \alpha u~\sum\limits_{k_1=1}^d\sum\limits_{k_2=1}^d~a_{k_1,k_2}~\Pi^{(k_1)}_{\bm{\mu}}\otimes \Pi^{(k_2)}_{\bm{\nu}}.$$

The Yang-Baxter algebras are studied. As a consequence the transfer matrix built out of 
$$
\check{R}_{j,j+1}(u;\mu,\nu) = \mathbb{1} +  u\left[\left(\Pi_{\mu\mu}\right)_{j}\left(\Pi_{\nu\nu}\right)_{j+1}\right]=\mathbb{1} + \frac{u}{(1+\mu^2)(1+\nu^2)}\begin{pmatrix}
        1 & \mu \\ \mu & \mu^2
    \end{pmatrix}_j \begin{pmatrix}
        1 & \nu \\ \nu & \nu^2
    \end{pmatrix}_{j+1}
$$
is commuting for different values of $u$ and $\mu$. A similar property holds for the non-additive $R$-matrices of \eqref{eq:generalRnonfactorised}. As a result the spectrum of the transfer matrices, \eqref{eq:transferMatrixeven}, \eqref{eq:transferMatrixodd}, are easily obtained (See Table \ref{tab:spectrumtransferMatrix}). 

The new $R$-matrices introduced here need new adaptations of the algebraic Bethe ansatz. Similar solutions also need to be obtained for the reflection equations.

\section*{Acknowledgments}
This research is funded by the U.S. Department of Energy, Office of Science, National Quantum Information Science Research Centers, Co-Design Center for Quantum Advantage under Contract No. DE-SC0012704.
The work of K. H. was supported by the National Natural Science Foundation
of China (Grant Nos. 12275214, 12247103, and 12047502), the Natural Science Basic Research Program of Shaanxi Province Grant Nos. 2021JCW-19 and 2019JQ-107, and Shaanxi Key Laboratory for Theoretical Physics Frontiers in China.


\appendix

\section{Non-additive $R$-matrices based on supercharges}
\label{app:SUSYnar}
To examine the scope of the $R$-matrices in \eqref{eq:generalRnonfactorised} we present solutions based on other examples of Abelian structures. Supersymmetry (SUSY) algebras provide such an example. Consider a supercharge acting on $\mathbb{C}^2$ depending on two parameters, $\mu$, $\nu$ : 
\begin{equation}\label{eq:qmunu}
    q(\mu, \nu) = \mathrm{i}\sqrt{\mu^2 + \nu^2}~X + \mu~Y + \nu~Z, 
\end{equation}
where $X$, $Y$ and $Z$ are the three Pauli matrices acting on $\mathbb{C}^2$. 
It is easy to verify that these supercharges are nilpotent, 
\begin{equation}
    q^2(\mu, \nu)=0.
\end{equation}
Then the $R$-matrix,
\begin{equation}\label{eq:Rsusy}
    R_{12}(\mu,\nu;\alpha,\beta) = 1 + q_1(\mu,\nu)q_2(\alpha,\beta),
\end{equation}
satisfies the YBE in the form,
\begin{equation}
    R_{12}(\mu,\nu;\alpha, \beta)R_{13}(\mu,\nu;\gamma, \delta)R_{23}(\alpha, \beta;\gamma, \delta) = R_{23}(\alpha, \beta;\gamma, \delta)R_{13}(\mu,\nu;\gamma, \delta)R_{12}(\mu,\nu;\alpha, \beta).
\end{equation}
As the SUSY Hamiltonians
\begin{equation}
    h(\mu,\nu) = \{q(\mu, \nu), q^\dag(\mu, \nu)\},
\end{equation}
commute with the supercharges, we can construct more non-additive solutions as
\begin{equation}\label{eq:Rsusy2}
    R_{12}(\mu,\nu;\alpha,\beta) = 1 + q_1(\mu,\nu)q_2(\alpha,\beta)+h_1(\mu,\nu)h_2(\alpha,\beta).
\end{equation}
These solutions can be extended to qudit spaces, $\mathbb{C}^d$. The supercharges in \eqref{eq:qmunu} will be appropriately modified and the resulting $R$-matrices will depend on more parameters.

\section{The inequivalence of \eqref{eq:projectorNARgeneralqubit} to known non-additive solutions}
\label{app:inequivalence}
Two $R$ matrices are equivalent if they are related to each other a local similarity transformation. More precisely we require
\begin{equation}\label{eq:Requivalence1}
    \tilde{R}(\mu,\nu) = \left[W(\mu)\otimes W(\nu)\right]R(\mu,\nu)\left[W^{-1}(\mu)\otimes W^{-1}(\nu)\right].
\end{equation}
It is easily seen that $\tilde{R}$ solves the YBE in \eqref{eq:qYBEnonadditive} if $R$ is a solution. Moreover
\begin{equation}\label{eq:Requivalence2}
    \tilde{R}(\mu, \nu) = g(\mu, \nu)R(\mu, \nu),
\end{equation}
implies that $\tilde{R}$ is a solution of \eqref{eq:YBEnonadditive} when $R$ is a solution. We will use these relations as the criteria to determine if two $R$-matrices fall in the same equivalence class. 

We will compare the most general 4 by 4 $R$-matrix in this paper, \eqref{eq:projectorNARgeneralqubit} with 4 by 4 non-additive $R$-matrices in \cite{Corcoran2023AllR}. The eigenvalues of the two $R$-matrices are compared to establish a similarity transform of the type shown in \eqref{eq:Requivalence1}. If the eigenvalues are not the same and have different multiplicities then we can be sure that no such similarity transform exists. The eigenvalues of the 4 by 4 non-additive $R$-matrix in \eqref{eq:projectorNARgeneralqubit} are given by,
\begin{equation}\label{eq:eigenvalueset}
    \{1+a_{1,1}, 1+a_{1,2}, 1+a_{2,1}, 1+a_{2,2}\}.
\end{equation}
Thus there are four different non-degenerate eigenvalues. The non-additive $R$-matrices denoted Models 1, 2, 3, and 4 in Section 4.1 of \cite{Corcoran2023AllR} have eigenvalue sets,
\begin{eqnarray}
    & \{-1,1,1,1\} & \\
    & \{-1,1,1,1\} & \\
    & \{-e^{F(u)-F(v)},e^{F(u)-F(v)},1,1\} & \\
    & \{1-\frac{2}{1+F(u)-F(v)},1,1,1\} &,
\end{eqnarray}
respectively. The first two sets can be mapped to the eigenvalues of the $R$-matrix in \eqref{eq:projectorNARgeneralqubit} by setting $a_{1,2}=a_{2,1}=a_{2,2}=0$ and $a_{1,1}=-2$. However a simple computation in Mathematica shows that no similarity transformation exists. The second two sets, corresponding to models 3 and 4 have eigenvalues that cannot be mapped to the eigenvalues in \eqref{eq:eigenvalueset}. 

Next we compare the eigenvalues of the three models denoted {\it 6vB, off-diagonal model, 8vB} in Section 4.3.2 of \cite{Corcoran2023AllR}. None of their eigenvalues can be mapped to the eigenvalues in \eqref{eq:eigenvalueset}. Thus we conclude that the non-additive $R$-matrix in \eqref{eq:projectorNARgeneralqubit} does not fall into the equivalence classes generated by the non-additive $R$-matrices of \cite{Corcoran2023AllR}.

Finally we check the equivalence with the non-additive $R$-matrix constructed in \eqref{eq:Rsusy}. The latter has a single eigenvalue 1 with multiplicity 4. However this $R$-matrix cannot be transformed to the 4 by 4 identity matrix using the transformation in \eqref{eq:Requivalence1}. This also implies that it is inequivalent to the non-additive $R$-matrix in \eqref{eq:projectorNARgeneralqubit}.

\section{Proof for the relations \eqref{relation_B-C}, \eqref{relation_A-B} and \eqref{relation_A-D}}
\label{app:relations_ABD}

The monodromy matrix is the product of a series of local monodromy matrices ($L$ operators, in our case $\check{R}$ matrices). They can also be expressed in the auxiliary space $0$,
\begin{eqnarray}
T(u;{\mu}, {\nu}) =
\begin{pmatrix} \mathcal{A}(u;{\mu}, {\nu}) & \mathcal{B}(u;{\mu}, {\nu}) \cr \mathcal{C}(u;{\mu}, {\nu}) & \mathcal{D}(u;{\mu}, {\nu}) \end{pmatrix}_0=
\begin{pmatrix} A_N & B_N \cr C_N & D_N \end{pmatrix}_0 \cdots
\begin{pmatrix} A_1 & B_1 \cr C_1 & D_1 \end{pmatrix}_0.
\end{eqnarray}
We prove the relations \eqref{relation_B-C}, \eqref{relation_A-B} and \eqref{relation_A-D} by induction in this Appendix. 

In the following proof procedure, like in section \ref{sec:YangBaxterAlgebra} of the main text, we denote the operators $\mathcal{A}(u;{\mu}, {\nu})$ , $\mathcal{B}(u;{\mu}, {\nu})$, $\mathcal{C}(u;{\mu}, {\nu})$, and $\mathcal{D}(u;{\mu}, {\nu})$ of total system sites $N$ as $\mathcal{A}_N$ , $\mathcal{B}_N$, $\mathcal{C}_N$, and $\mathcal{D}_N$, respectively. The operators $A$, $B$, $C$, and $D$ in regular typeface with subscripts are local operators
acting on the quantum spaces from $1$ till $N$ .

For the total system sites $N=1$,
\begin{eqnarray}
T(u;{\mu}, {\nu}) =
\begin{pmatrix} \mathcal{A}_1 & \mathcal{B}_1 \cr \mathcal{C}_1 & \mathcal{D}_1 \end{pmatrix}_0 
=
\begin{pmatrix} A_1 & B_1 \cr C_1 & D_1 \end{pmatrix}_0=
\begin{pmatrix}
\mathbb{1}+\frac{u}{1+\mu^2} (\Pi_{\nu\nu})_1 & \frac{u\mu}{1+\mu^2} (\Pi_{\nu\nu})_1 \cr
\frac{u\mu}{1+\mu^2} (\Pi_{\nu\nu})_1 & \mathbb{1}+\frac{u\mu^2}{1+\mu^2} (\Pi_{\nu\nu})_1
\end{pmatrix}.
\end{eqnarray}
Here $\mathcal{A}_1=A_1=\mathbb{1}+\frac{u}{1+\mu^2} (\Pi_{\nu\nu})_1$, $\mathcal{B}_1=\mathcal{C}_1=B_1=\frac{u\mu}{1+\mu^2} (\Pi_{\nu\nu})_1$, and $\mathcal{D}_1=D_1=\mathbb{1}+\frac{u\mu^2}{1+\mu^2} (\Pi_{\nu\nu})_1$ .
It is obvious that 
\begin{eqnarray}
\mathcal{A}_1 \mu-\mathcal{B}_1=\mu\mathbb{1}, \quad\text{and}\quad 
\mathcal{A}_1 \mu^2-\mathcal{D}_1=(\mu^2-1)\mathbb{1}.
\end{eqnarray}

For the system sites $N=2$ case,
\begin{eqnarray}
T(u;{\mu}, {\nu})& = &
\begin{pmatrix} \mathcal{A}_2 & \mathcal{B}_2 \cr \mathcal{C}_2 & \mathcal{D}_2 \end{pmatrix}
=\begin{pmatrix} A_2 & B_2 \cr C_2 & D_2 \end{pmatrix}
\begin{pmatrix} A_1 & B_1 \cr C_1 & D_1 \end{pmatrix}
\nonumber\\
& = &
\begin{pmatrix}
\mathbb{1}+\frac{u}{1+\mu^2} (\Pi_{\nu\nu})_2 & \frac{u\mu}{1+\mu^2} (\Pi_{\nu\nu})_2 \cr
\frac{u\mu}{1+\mu^2} (\Pi_{\nu\nu})_2 & \mathbb{1}+\frac{u\mu^2}{1+\mu^2} (\Pi_{\nu\nu})_2
\end{pmatrix}
\begin{pmatrix}
\mathbb{1}+\frac{u}{1+\mu^2} (\Pi_{\nu\nu})_1 & \frac{u\mu}{1+\mu^2} (\Pi_{\nu\nu})_1 \cr
\frac{u\mu}{1+\mu^2} (\Pi_{\nu\nu})_1 & \mathbb{1}+\frac{u\mu^2}{1+\mu^2} (\Pi_{\nu\nu})_1
\end{pmatrix}.
\end{eqnarray}
We can write down the expressions for $\mathcal{A}_2$, $\mathcal{B}_2$, $\mathcal{C}_2$, and $\mathcal{D}_2$ by matrix multiplication,
\begin{eqnarray}
\mathcal{A}_2&=&\left(\mathbb{1}+\frac{u}{1+\mu^2}(\Pi_{\nu\nu})_1\right)\left(\mathbb{1}+\frac{u}{1+\mu^2}(\Pi_{\nu\nu})_2\right)
+\left(\frac{u\mu}{1+\mu^2}\right)^2(\Pi_{\nu\nu})_1(\Pi_{\nu\nu})_2,\\
\mathcal{B}_2&=&\left(\mathbb{1}+\frac{u}{1+\mu^2}(\Pi_{\nu\nu})_2\right)\frac{u\mu}{1+\mu^2}(\Pi_{\nu\nu})_1+
\left(\mathbb{1}+\frac{u\mu^2}{1+\mu^2}(\Pi_{\nu\nu})_1\right)\frac{u\mu}{1+\mu^2}(\Pi_{\nu\nu})_2,\\
\mathcal{C}_2&=&\left(\mathbb{1}+\frac{u}{1+\mu^2}(\Pi_{\nu\nu})_1\right)\frac{u\mu}{1+\mu^2}(\Pi_{\nu\nu})_2+
\left(\mathbb{1}+\frac{u\mu^2}{1+\mu^2}(\Pi_{\nu\nu})_2\right)\frac{u\mu}{1+\mu^2}(\Pi_{\nu\nu})_1,\\
\mathcal{D}_2&=&\left(\mathbb{1}+\frac{u\mu^2}{1+\mu^2}(\Pi_{\nu\nu})_1\right)\left(\mathbb{1}+\frac{u\mu^2}{1+\mu^2}(\Pi_{\nu\nu})_2\right)
+\left(\frac{u\mu}{1+\mu^2}\right)^2(\Pi_{\nu\nu})_1(\Pi_{\nu\nu})_2.
\end{eqnarray}
It is easy to check that $\mathcal{C}_2=\mathcal{B}_2$. Expand the expression for $\mathcal{A}_2$ and multiply it by $\mu^2$, then compare the result with the expansion for $\mathcal{D}_2$
\begin{eqnarray}
\mathcal{A}_2\mu^2&=&\mathbb{1}\mu^2+\frac{u\mu^2}{1+\mu^2}(\Pi_{\nu\nu})_1+\frac{u\mu^2}{1+\mu^2}(\Pi_{\nu\nu})_2\nonumber\\
&&+\left(\frac{u\mu}{1+\mu^2}\right)^2(\Pi_{\nu\nu})_1(\Pi_{\nu\nu})_2+\mu^2\left(\frac{u\mu}{1+\mu^2}\right)^2(\Pi_{\nu\nu})_1(\Pi_{\nu\nu})_2,\\
\mathcal{D}_2&=&\mathbb{1}+\frac{u\mu^2}{1+\mu^2}(\Pi_{\nu\nu})_1+\frac{u\mu^2}{1+\mu^2}(\Pi_{\nu\nu})_2\nonumber\\
&&+\left(\frac{u\mu^2}{1+\mu^2}\right)^2(\Pi_{\nu\nu})_1(\Pi_{\nu\nu})_2+\left(\frac{u\mu}{1+\mu^2}\right)^2(\Pi_{\nu\nu})_1(\Pi_{\nu\nu})_2.
\end{eqnarray}
Only the first terms with identity operators in the above two equations are different, so we have
\begin{eqnarray}
\mathcal{A}_2\mu^2-\mathcal{D}_2=(\mu^2-1)\mathbb{1}.
\end{eqnarray}
Similarly, multiply the expansion for $\mathcal{A}_2$ by $\mu$, and compare it with the expansion for $\mathcal{B}_2$
\begin{eqnarray}
\mathcal{A}_2\mu&=&\mathbb{1}\mu+\frac{u\mu}{1+\mu^2}(\Pi_{\nu\nu})_1+\frac{u\mu}{1+\mu^2}(\Pi_{\nu\nu})_2\nonumber\\
&&+\mu\left(\frac{u}{1+\mu^2}\right)^2(\Pi_{\nu\nu})_1(\Pi_{\nu\nu})_2+\mu\left(\frac{u\mu}{1+\mu^2}\right)^2(\Pi_{\nu\nu})_1(\Pi_{\nu\nu})_2 ,\\
\mathcal{B}_2&=&\frac{u\mu}{1+\mu^2}(\Pi_{\nu\nu})_1+\frac{u\mu}{1+\mu^2}(\Pi_{\nu\nu})_2\nonumber\\
&&+\mu\left(\frac{u}{1+\mu^2}\right)^2(\Pi_{\nu\nu})_1(\Pi_{\nu\nu})_2+\mu\left(\frac{u\mu}{1+\mu^2}\right)^2(\Pi_{\nu\nu})_1(\Pi_{\nu\nu})_2 .
\end{eqnarray}
Only the expansion for $\mathcal{A}_2\mu$ has identity term, thus
\begin{eqnarray}
\mathcal{A}_2\mu-\mathcal{B}_2=\mu\mathbb{1}.
\end{eqnarray}
Assume the relations \eqref{relation_B-C}, \eqref{relation_A-B} and \eqref{relation_A-D} hold for the system sites $N-1$ case, then for the system with sites $N$,
\begin{eqnarray}
T(u;{\mu}, {\nu})& = & 
\begin{pmatrix} \mathcal{A}_N & \mathcal{B}_N \cr \mathcal{C}_N & \mathcal{D}_N \end{pmatrix} 
=\begin{pmatrix} A_N & B_N \cr C_N & D_N \end{pmatrix}
\begin{pmatrix} \mathcal{A}_{N-1} & \mathcal{B}_{N-1} \cr \mathcal{C}_{N-1} & \mathcal{D}_{N-1} \end{pmatrix} \cr
& = &
\begin{pmatrix} A_N\mathcal{A}_{N-1}+B_N\mathcal{C}_{N-1} & A_N\mathcal{B}_{N-1}+B_N\mathcal{D}_{N-1} \cr
C_N\mathcal{A}_{N-1}+D_N\mathcal{C}_{N-1} & C_N\mathcal{B}_{N-1}+D_N\mathcal{D}_{N-1} \end{pmatrix}.
\end{eqnarray}
We have the expressions for $\mathcal{A}_N$, $\mathcal{B}_N$, $\mathcal{C}_N$, and $\mathcal{D}_N$ in terms of $\mathcal{A}_{N-1}$, $\mathcal{B}_{N-1}$, $\mathcal{C}_{N-1}$, and $\mathcal{D}_{N-1}$
\begin{eqnarray}
\mathcal{A}_{N}=\left(\mathbb{1}+\frac{u}{1+\mu^2}(\Pi_{\nu\nu})_N\right)\mathcal{A}_{N-1}+\frac{u\mu}{1+\mu^2}(\Pi_{\nu\nu})_N\mathcal{C}_{N-1},\label{eq:recursive-A}\\
\mathcal{B}_{N}=\left(\mathbb{1}+\frac{u}{1+\mu^2}(\Pi_{\nu\nu})_N\right)\mathcal{B}_{N-1}+\frac{u\mu}{1+\mu^2}(\Pi_{\nu\nu})_N\mathcal{D}_{N-1},\label{eq:recursive-B}\\
\mathcal{C}_{N}=\frac{u\mu}{1+\mu^2}(\Pi_{\nu\nu})_N\mathcal{A}_{N-1}+\left(\mathbb{1}+\frac{u\mu^2}{1+\mu^2}(\Pi_{\nu\nu})_N\right)\mathcal{C}_{N-1},\label{eq:recursive-C}\\
\mathcal{D}_{N}=\frac{u\mu}{1+\mu^2}(\Pi_{\nu\nu})_N\mathcal{B}_{N-1}+\left(\mathbb{1}+\frac{u\mu^2}{1+\mu^2}(\Pi_{\nu\nu})_N\right)\mathcal{D}_{N-1}.\label{eq:recursive-D}
\end{eqnarray}
Both the expressions for $\mathcal{A}_{N}$ and $\mathcal{C}_{N}$ are composed of $\mathcal{A}_{N-1}$ and $\mathcal{C}_{N-1}$, we have that
\begin{eqnarray}
\mathcal{A}_{N}\mu-\mathcal{C}_{N}=\mathcal{A}_{N-1}\mu-\mathcal{C}_{N-1}=\mu\mathbb{1}.
\end{eqnarray}
We can calculate the difference between $\mathcal{B}_{N}$ and $\mathcal{C}_{N}$, 
\begin{eqnarray}
\mathcal{B}_{N}-\mathcal{C}_{N}&=&\left(\mathbb{1}+\frac{u}{1+\mu^2}(\Pi_{\nu\nu})_N\right)\mathcal{B}_{N-1}-\left(\mathbb{1}+\frac{u\mu^2}{1+\mu^2}(\Pi_{\nu\nu})_N\right)\mathcal{C}_{N-1}\nonumber\\
&&+\frac{u\mu}{1+\mu^2}(\Pi_{\nu\nu})_N\mathcal{D}_{N-1}-\frac{u\mu}{1+\mu^2}(\Pi_{\nu\nu})_N\mathcal{A}_{N-1}\nonumber\\
&=&\frac{u-u\mu^2}{1+\mu^2}(\Pi_{\nu\nu})_N\mathcal{B}_{N-1}
+\frac{u\mu}{1+\mu^2}(\Pi_{\nu\nu})_N\big(\mathcal{D}_{N-1}-\mathcal{A}_{N-1}\big)\nonumber\\
&=&\frac{u}{1+\mu^2}(\Pi_{\nu\nu})_N
\left((1-\mu^2)\mathcal{B}_{N-1}
+\mu(\mathcal{A}_{N-1}\mu^2+(1-\mu^2)-\mathcal{A}_{N-1})
\right)\nonumber\\
&=&\frac{u}{1+\mu^2}(\Pi_{\nu\nu})_N
\left((1-\mu^2)(\mathcal{A}_{N-1}\mu-\mu)
+\mu(\mathcal{A}_{N-1}\mu^2+(1-\mu^2)-\mathcal{A}_{N-1})
\right)\nonumber\\
&=&0 .
\end{eqnarray}
In the above derivation, we have taken into account that $\mathcal{B}_{N-1}=\mathcal{C}_{N-1}$, and $\mathcal{B}_{N-1}$ and $\mathcal{D}_{N-1}$ can be expressed in terms of $\mathcal{A}_{N-1}$.

We proceed to calculate $\mathcal{A}_{N}\mu^2-\mathcal{D}_{N}$
\begin{eqnarray}
\mathcal{A}_{N}\mu^2-\mathcal{D}_{N}&=&\mathcal{A}_{N-1}\mu^2+\mathcal{A}_{N-1}\frac{u\mu^2}{1+\mu^2}(\Pi_{\nu\nu})_N
+\mathcal{C}_{N-1}\frac{u\mu^3}{1+\mu^2}(\Pi_{\nu\nu})_N\nonumber\\
&&-\mathcal{B}_{N-1}\frac{u\mu}{1+\mu^2}(\Pi_{\nu\nu})_N-\mathcal{D}_{N-1}\left(\mathbb{1}+\frac{u\mu^2}{1+\mu^2}(\Pi_{\nu\nu})_N\right)\nonumber\\
&=&\mathcal{A}_{N-1}\mu^2-\mathcal{D}_{N-1}+(\mathcal{A}_{N-1}-\mathcal{D}_{N-1})\frac{u\mu^2}{1+\mu^2}(\Pi_{\nu\nu})_N
+\mathcal{B}_{N-1}\frac{u\mu}{1+\mu^2}(\Pi_{\nu\nu})_N(\mu^2-1)\nonumber\\
&=&\mathcal{A}_{N-1}\mu^2-\mathcal{D}_{N-1}+\left(\mathcal{A}_{N-1}\mu-\mathcal{D}_{N-1}\mu+\mathcal{B}_{N-1}(\mu^2-1)\right)\frac{u\mu}{1+\mu^2}(\Pi_{\nu\nu})_N
.\label{eq:AN-DN}\end{eqnarray}
Substitute the following two relations for $N-1$ to the above equation
\begin{eqnarray}
\mathcal{D}_{N-1}&=&\mathcal{A}_{N-1}\mu^2+\mathbb{1}-\mu^2,\\
\mathcal{B}_{N-1}&=&\mathcal{A}_{N-1}\mu-\mu.
\end{eqnarray}
We obtain that the part in the round brackets in the last line of \eqref{eq:AN-DN} is equal to $0$. So we have that
\begin{eqnarray}
\mathcal{A}_{N}\mu^2-\mathcal{D}_{N}=\mathcal{A}_{N-1}\mu^2-\mathcal{D}_{N-1}=(\mu^2 - 1) \mathbb{1}.
\end{eqnarray}
The relations \eqref{relation_B-C}, \eqref{relation_A-B} and \eqref{relation_A-D} are proved.

\section{Properties of the transfer matrices \eqref{eq:transferMatrixeven} and \eqref{eq:transferMatrixodd}}
\label{app:transfermatrices}
We derive the expressions for the transfer matrices when the length of the spin chain is odd or even. To do this we note that the $R$-matrix, \eqref{eq:R-matrix} can be equivalently written as
\begin{equation}
    \check{R}_{0,j}(u;\mu,\nu) = \begin{pmatrix}
        \mathbb{1}+l\left(\Pi_{\nu\nu}\right)_j & \mu l \left(\Pi_{\nu\nu}\right)_j \\ \mu l \left(\Pi_{\nu\nu}\right)_j & \mathbb{1} + \mu^2l\left(\Pi_{\nu\nu}\right)_j
    \end{pmatrix} = \alpha_j \mathbb{1} + \beta\left(\Pi_{\nu\nu}\right)_j Z + \gamma \left(\Pi_{\nu\nu}\right)_j X.
\end{equation}
Here $X$ and $Z$ are the first and third Pauli matrices respectively and $\mathbb{1}$ is the 2 by 2 identity operator. They act on the auxiliary space indexed by 0. The parameters $\alpha_j$, $\beta$ and $\gamma$ are
\begin{equation}
    \alpha_j  =  \mathbb{1} + \frac{u}{2}\left(\Pi_{\nu\nu}\right)_j,~~
    \beta  = \frac{l(1-\mu^2)}{2},~~\gamma = \mu l,~;~l=\frac{u}{1+\mu^2}.
\end{equation}
With this expression the monodromy matrix reads
\begin{equation}
    T(u;\mu,\nu) = \prod\limits_{j=1}^N~\left[\alpha_j \mathbb{1} + \beta\left(\Pi_{\nu\nu}\right)_j Z + \gamma \left(\Pi_{\nu\nu}\right)_j X\right].
\end{equation}
This product is a linear combination of the 2 by 2 matrices, $\mathbb{1}$, $X$ and $Z$. The coefficients of this linear combination are the operators $\left(\Pi_{\nu\nu}\right)_j$  acting on the physical Hilbert space. As a result the trace, and hence the transfer matrix, is precisely the coefficient of the identity operator as the $X$ and $Z$ operators are traceless. Thus obtaining the transfer matrix amounts to identifying all possible ways to get the identity operator in this product. We need to identify the even powers of $X$ and $Z$ in this product as $X^2=\mathbb{1}$ and $Z^2=\mathbb{1}$. For example when $N=2$ the product becomes,
\begin{eqnarray}
    \prod\limits_{j=1}^2~\left[\alpha_j \mathbb{1} + \beta\left(\Pi_{\nu\nu}\right)_j Z + \gamma \left(\Pi_{\nu\nu}\right)_j X\right] & = & \left[\alpha_1\alpha_2 +\left(\beta^2+\gamma^2\right)\left(\Pi_{\nu\nu}\right)_1\left(\Pi_{\nu\nu}\right)_2\right] \mathbb{1} \nonumber \\
    & + & \textrm{traceless terms}.
\end{eqnarray}
This yields the transfer matrix,
\begin{equation}
    \mathcal{T}(u;\mu,\nu) = 2\left[\alpha_1\alpha_2 +\left(\beta^2+\gamma^2\right)\left(\Pi_{\nu\nu}\right)_1\left(\Pi_{\nu\nu}\right)_2\right],
\end{equation}
where we have used $\Tr{\mathbb{1}}=2$. When $N=3$ we find,
\begin{eqnarray}
    \prod\limits_{j=1}^3~\left[\alpha_j \mathbb{1} + \beta\left(\Pi_{\nu\nu}\right)_j Z + \gamma \left(\Pi_{\nu\nu}\right)_j X\right] & = & \left[\alpha_1\alpha_2\alpha_3 \right. \nonumber \\ &  + & \left. \alpha_1\left(\beta^2+\gamma^2\right)\left(\Pi_{\nu\nu}\right)_1\left(\Pi_{\nu\nu}\right)_2 \right. \nonumber \\ & + & \left.   \alpha_2\left(\beta^2+\gamma^2\right)\left(\Pi_{\nu\nu}\right)_1\left(\Pi_{\nu\nu}\right)_3 \right. \nonumber \\ & + & \left. \alpha_3\left(\beta^2+\gamma^2\right)\left(\Pi_{\nu\nu}\right)_2\left(\Pi_{\nu\nu}\right)_3\right] \mathbb{1} \nonumber \\
    & + & \textrm{traceless terms}.
\end{eqnarray}
These two examples show that the product will depend on whether the length of the chain, $N$, is odd or even. The above examples imply that we need to pick powers of $\left(\beta^2 + \gamma^2\right)$ from the product to obtain a coefficient of the 2 by 2 identity operator. There are a total of $\binom{N}{2k}$ ways of doing this. It results in terms of the form
$$ \left(\beta^2 + \gamma^2\right)^k\mathop{\prod\limits_{j=1}^N}_{j\neq\{i_1,\cdots, i_{2k} \}}\alpha_j\left(\Pi_{\nu\nu}\right)_{i_1}\cdots\left(\Pi_{\nu\nu}\right)_{i_{2k}}. $$
We repeat this step for $k=1$ up to $k=\lfloor\frac{N}{2}\rfloor$. Following this procedure we arrive at the expressions in \eqref{eq:transferMatrixeven}
and \eqref{eq:transferMatrixodd}.

Next we systematically construct the eigenstates of the transfer matrix for periodic spin chains of length $N$. The latter can be odd or even. We start with the product state, $$ \ket{\left(0_\nu\right)_1\cdots \left(0_\nu\right)_N}.$$ The transfer matrix acts on this state through the term $\prod\limits_{j=1}^N~\alpha_j$ in both the even and odd cases. The remaining terms in the sums \eqref{eq:transferMatrixeven} and \eqref{eq:transferMatrixodd} give 0. We also have 
$$ \alpha_j\ket{\left(0_\nu\right)_j} = \ket{\left(0_\nu\right)_j}.$$
Thus the eigenvalue of this state is 2. We will consider this as a reference state to construct the rest of the spectrum. Next we consider states where one of the $0_\nu$'s from the reference state is flipped to a $1_\nu$ state $$ \left\{\ket{\left(0_\nu\right)_1\cdots\left(1_\nu\right)_j\cdots\left(0_\nu\right)_N}~|~j\in\{1,\cdots, N \}\right\}.$$ 
There are $N$ of them and we will call them the {\it `1-particle' states}. When $N$ is even, the only term with a non-zero action on this state is the $\prod\limits_{j=1}^{N}~\alpha_j$ term. The corresponding eigenvalue is $2\left(1 + \frac{u}{2}\right)$. It is easily seen that even when $N$ is odd just the $k=N$ term has a non-zero action on the state. Thus the eigenvalue is once again, $2\left(1 + \frac{u}{2}\right)$. Next we consider states where exactly two of the $0_\nu$'s are flipped to $1_\nu$'s,
$$ \left\{\ket{\left(0_\nu\right)_1\cdots\left(1_\nu\right)_{j_1}\cdots\left(1_\nu\right)_{j_2}\cdots\left(0_\nu\right)_N}~|~j_1,j_2\in\{1,\cdots, N \}\right\}.$$
There are $\binom{N}{2}$ such {\it `2-particle states'} states. We find that for both even and odd values of $N$ the eigenvalue is $$ 2\left(1 + \frac{u}{2}\right)^2 + 2\left(\beta^2  + \gamma^2\right)^2.$$
These two cases suggest that the difference in eigenvalues depends on whether the number of $1_\nu$'s in a given product state is odd or even. Thus the eigenvalues for a general {\it `$2f+1$-particle state'} is given by
\begin{equation}
    2\left(1 + \frac{u}{2}\right)^{2f+1},
\end{equation}
and for a general {\it `$2f$-particle state'} is given by,    
\begin{equation}
    2\left(1 + \frac{u}{2}\right)^{2f} + 2\left(\beta^2  + \gamma^2\right)^f.
\end{equation}
The degeneracies for the odd and even cases are $\binom{N}{2f+1}$ and $\binom{N}{2f}$ respectively Note that $$ \beta^2 + \gamma^2 = \frac{u^2}{4}.$$ This makes all the eigenvalues of the transfer matrix dependent only on the spectral parameter $u$ and independent of $\mu$ and $\nu$.

Finally we consider the state
$$ \ket{\left(1_\nu\right)_1\cdots \left(1_\nu\right)_N},$$
where all the $0_\nu$'s are flipped to $1_\nu$. All the terms in the expression for the transfer matrices, \eqref{eq:transferMatrixeven} and \eqref{eq:transferMatrixodd} have an action on this state. Consequently the eigenvalues are found to be 
$$ 2\sum\limits_{k=0}^N~\binom{2N+1}{2k+1}\left(1 + \frac{u}{2}\right)^{2k+1}\left(\frac{u}{2}\right)^{\left[(2N+1)-(2k+1)\right]},$$
and 
$$ 2\sum\limits_{k=0}^N~\binom{2N}{2k}\left(1 + \frac{u}{2}\right)^{2k}\left(\frac{u}{2}\right)^{\left(2N-2k\right)},$$
for the odd and even cases respectively. 


\bibliographystyle{unsrt}
\normalem
\bibliography{refs}

\end{document}